\documentclass{sf2a-conf2011}
\usepackage{graphicx}
\usepackage{hyperref}
\usepackage[]{natbib}
\usepackage[cyr]{aeguill}
\usepackage{epstopdf}

\def\BibTeX{{\rm B\kern-.05em{\sc i\kern-.025em b}\kern-.08em
    T\kern-.1667em\lower.7ex\hbox{E}\kern-.125emX}}
\bibpunct{(}{)}{;}{a}{}{,}  %%%%%%%%%%%%%  A&A bibliography style

\begin{document}

\TitreGlobal{SF2A 2011}

\title{Accurate Black Hole Mass Measurements for Thermal AGNs and the Origin of the Correlations Between Black Hole Mass and Bulge Properties}

\runningtitle{AGN black hole masses}

\author{C. M. Gaskell}\address{Centro de Astrof\'isica de Valpara\'iso y Departamento de F\'isica y Astronom\'ia,  Universidad de Valpara\'iso, Av. Gran Breta\~na 1111, Valpara\'iso, Chile}

\setcounter{page}{237}

\index{Gaskell, C.\,M.}

%%-----------------------------------------------------------------

\maketitle

\begin{abstract}
A simple refinement is proposed to the Dibai method for determining black hole masses in type-1 thermal AGNs.  Comparisons with reverberation mapping black hole masses and host galaxy bulge properties suggest that the method is accurate to $\pm 0.15$ dex.  Contrary to what was thought when the $M_\bullet$ -- $\sigma_*$ relationship was first discovered, it does not have a lower dispersion than the $M_\bullet$ -- $L_{bulge}$ relationship.  The dispersion in the $M_\bullet$ -- $L_{host}$ relationship for AGNs decreases strongly with increasing  $M_\bullet$ or $L_{bulge}$.  This is naturally explained as a consequence of the $M_\bullet$ -- bulge relationships being the result of averaging due to mergers. Simulations show that the decrease in dispersion in the $M_\bullet$ or $L_{bulge}$ relationship with increasing mass is in qualitative agreement with being driven by mergers. The large scatter in AGN black hole masses at lower masses rules out significant AGN feedback.  A non-causal origin of the correlations between black holes and bulges explains the frequent lack of supermassive black holes in late-type galaxies, and the lack of correlation of black hole mass with pseudo-bulges.
\end{abstract}

%% Insert the keywords (to appear in the ADS indexing)
%% Keywords must be separated by a comma

\section{Introduction}

\citet{dibai77} introduced the estimation of the masses, $M_\bullet$, of supermassive black holes (SMBHs) in type-1 thermal AGNs (i.e., high accretion rate AGNs seen close to pole-on -- see \citealt{antonucci93,antonucci11}) from the optical luminosity and the FWHM of the H$\beta$ line.  Because of its ease of use, the ``Dibai method'' (also known as the ``photoionization method'' or the ``single-epoch method'') is by far the most widely used method of SMBH mass determination.  The assumptions in the Dibai method are discussed in \citet{bochkarev+gaskell09}.  For a recent review of AGN SMBH determinations see \citet{marziani+sulentic11}.  The Dibai method assumes that the broad-line region (BLR) is gravitationally bound and that AGNs have similar continuum shapes and structures so that the size of the H$\beta$ emitting region can be estimated from the luminosity.  For checking these assumptions and calibrating the method, reverberation mapping has been crucial.  Cross-correlation of line and continuum time series \citep{lyutyi+cherepashchuk72,cherepashchuk+lyutyi73,gaskell+sparke86} yields effective radii of the line-emitting regions, and velocity-resolved reverberation mapping shows that the gas is gravitationally bound \citep{gaskell88}.

A long-standing question in AGN research has been whether the masses of SMBHs are correlated with the masses of their host galaxies.  Until the mid-1990s it was generally assumed that an SMBH was an incidental and perhaps accidental addition to a galaxy -- discovering an SMBH in a galaxy was considered worthy of a press release!  A notable exception to this assumption was \citet{zasov+dibai70} who discovered a correlation between the brightness of AGNs and their host galaxies. \citet{dibai77} subsequently found an apparent correlation between AGN luminosity and $M_\bullet$.  Taken together these findings implied that $M_\bullet$ and $L_{host}$ were correlated.  The tacit assumption of the 1980s, though, was that any correlation between $M_\bullet$ and $L_{host}$ was the result of selection effects (because only a bright host could be seen in a bright AGN).  This attitude began to change when \citet{kormendy94} discovered that the masses of inactive SMBHs seemed to be correlated with the luminosity, $L_{bulge}$, of the bulge of the host galaxy.  Further work (notably \citealt{magorrian+98}) verified that there was indeed a $M_\bullet$ -- $L_{bulge}$ relationship\footnote{The correlation is with {\em bulge} luminosity \citep{kormendy+11}, but the $L_{host}$ measured by \citet{zasov+dibai70} would have been dominated by $L_{bulge}$.} but also showed that there was substantial scatter in it.  In trying to understand the origin of the relationship it is important to know how much of this scatter is the result of observational and modeling errors, and how much is intrinsic.

\section{Improving the Dibai method}
%%-------------------------

Masses determined from reverberation mapping \citep{gaskell88} have been considered the ``gold standard'' of AGN BH mass determinations because the effective radius is estimated relatively directly.   Comparing reverberation mapping masses, $M_{rev}$, with masses from the most widely-used version of the Dibai method ($M_{Dibai} \propto$ FWHM$^2 L^{0.5}$) gives a dispersion of $\pm 0.37$ dex (see Fig.\@ 1a).  It is known that BLR line profile shapes vary from object to object and that the shape is a function of line width (see \citealt{gaskell09b} for a review of the BLR).  It is therefore reasonable to ask whether the Dibai method can be refined.  \citet{gaskell10a} has suggested improving the Dibai method by introducing an extra line width term.  Fig.\@ 1b shows $M_{new}$, the masses determined from Eq.\@ 1 of \citet{gaskell10a}, compared with reverberation masses.  This gives a relative dispersion of $\pm 0.22$ dex.  An $F$-test shows that the improvement is significant at the 99.8\% level.

\vspace{0.2 cm}

%% Figure 1
\begin{figure}[ht!]
 \centering
 \includegraphics[width=0.8\textwidth,clip]{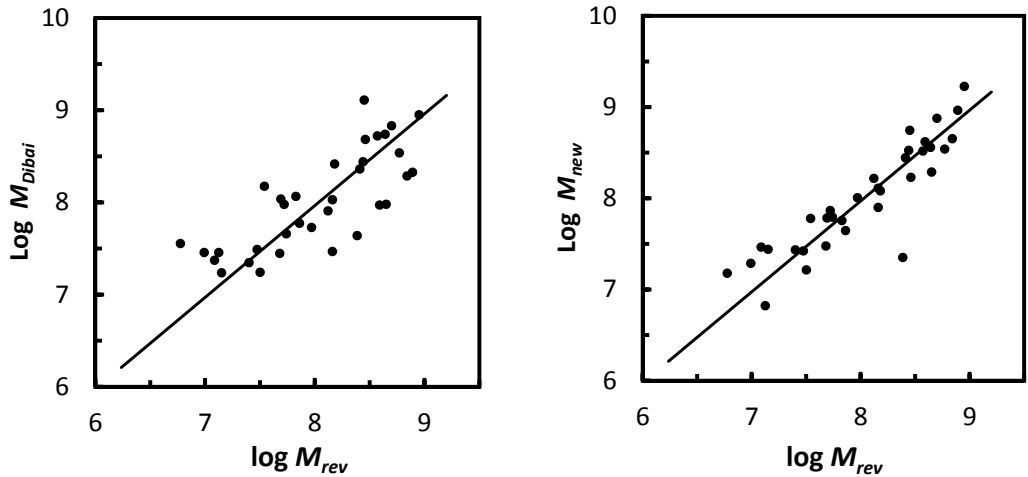}
%% Note the ABSENCE of the extension .pdf , .eps or .ps  !
  \caption{The left frame shows the correlation between masses, $M_{rev}$, measured from reverberation mapping and masses, $M_{Dibai}$, estimated using from FWHM$^2 L_{AGN}^{0.5}$.  The right frame shows the improvement when $M_{Dibai}$ is replaced with $M_{new}$, the mass estimated using Eq.\@ 1 of \citet{gaskell10a}.  H$\beta$ line widths, $L_{AGN}$, and $M_{rev}$ are from the compilation of \citet{vestergaard+peterson06}.}
  \label{Gaskell:Fig1}
\end{figure}

The $\pm 0.22$ dex includes errors from both reverberation mapping and from using Eq.\@ 1 of \citet{gaskell10a}.  From inspection of the \citet{vestergaard+peterson06} compilation of reverberation-mapping black hole estimates, the rms error in $M_{rev}$ is $\pm 0.12$ dex due to measuring error alone.  To this should be added an additional, unknown error caused by the off-axis nature of the continuum variability (see \citealt{gaskell10b}).  Because of this, reverberation mapping gives different time delays for different continuum events.  The effect of this on the error budget for $M_{rev}$ still needs to be evaluated.

\cite{denney+09} have used repeated measurements of the same object to estimate the error in the Dibai method due to measuring error.  They get an error of $\pm 0.10$ dex and, taking into account additional  effects, they believe the combined error is $\pm 0.12$ to 0.16 dex.  The rms error in $M_{new}$ will be similar.  These error estimates do not take into account possible random or systematic object-to-object differences in the scaling.

Since there could be unknown random or systematic sources of error in determination of $M_\bullet$ by each method, and some of these (such as orientation effects) could be the same for both $M_{rev}$ and $M_{new}$, it is important to have an external check.  The correlations between $M_\bullet$, $L_{bulge}$, and the stellar velocity dispersion, $\sigma_*$,  offer such a check.

\section{The $M_\bullet$ -- $L_{host}$ relationship for AGNs}

\citet{bentz+09} have presented host galaxy bulge luminosities for most of the AGNs with high-quality reverberation mapping mass estimates.  Fig.\@ 2a shows the host luminosities compared with masses estimated from FWHM$^2 L^{0.5}$.  Fig.\@ 2b shows the $M_\bullet$ -- $L_{bulge}$ relationship when the masses are estimated using Eq.\@ 1 of \citet{gaskell10a}.  As is expected from Fig.\@ 1, the scatter in the $M_\bullet$ -- $L_{bulge}$ relationship is reduced.  The interesting thing is that {\em it is only the scatter at the high-mass end that is reduced.}

\vspace{0.2 cm}

%% Figure 2
\begin{figure}[ht!]
 \centering
 \includegraphics[width=0.8\textwidth,clip]{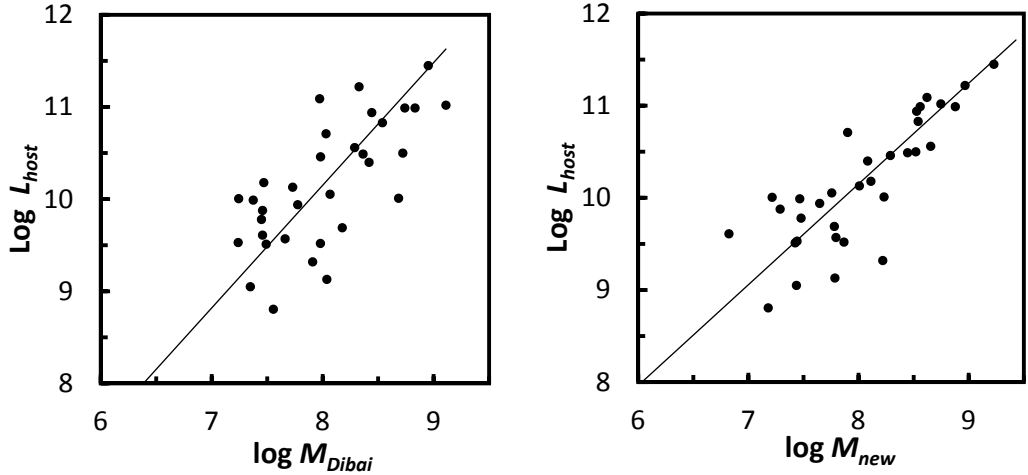}
%% Note the ABSENCE of the extension .pdf , .eps or .ps  !
  \caption{The improvement in the AGN $M_\bullet$ -- $L_{host}$ relationship when going from using the standard $M_{Dibai} \propto$ FWHM$^2 L^{0.5}$ masses (left-hand frame) to the masses, $M_{new}$, of \citet{gaskell10a} (right-hand frame).}
  \label{Gaskell:Fig2}
\end{figure}

Fig.\@ 3a shows the comparison of residuals from the $M_\bullet$ -- $L_{bulge}$ and $M_\bullet$ -- $\sigma_*$ relationships for the AGNs with $\sigma_*$ available.  The first thing to note is that the scatter in both axes is comparable.  Given that the excitement over the discovery of the $M_\bullet$ -- $\sigma_*$ relationship was that it was supposed to be significantly tighter than the $M_\bullet$ -- $L_{bulge}$ relationship (see Fig.\@ 2 of \citealt{gebhardt+00}), this is perhaps surprising, but making a similar plot for the compilation of non-BLR mass determinations in \citet{gultekin+09} (see Fig.\@ 3b) shows that non-BLR galaxies now have comparable scatter about the $M_\bullet$ -- $L_{bulge}$ and $M_\bullet$ -- $\sigma_*$ relationships too.

\vspace{0.2 cm}

%% Figure 3
\begin{figure}[ht!]
 \centering
 \includegraphics[width=0.35\textwidth,clip,angle=270]{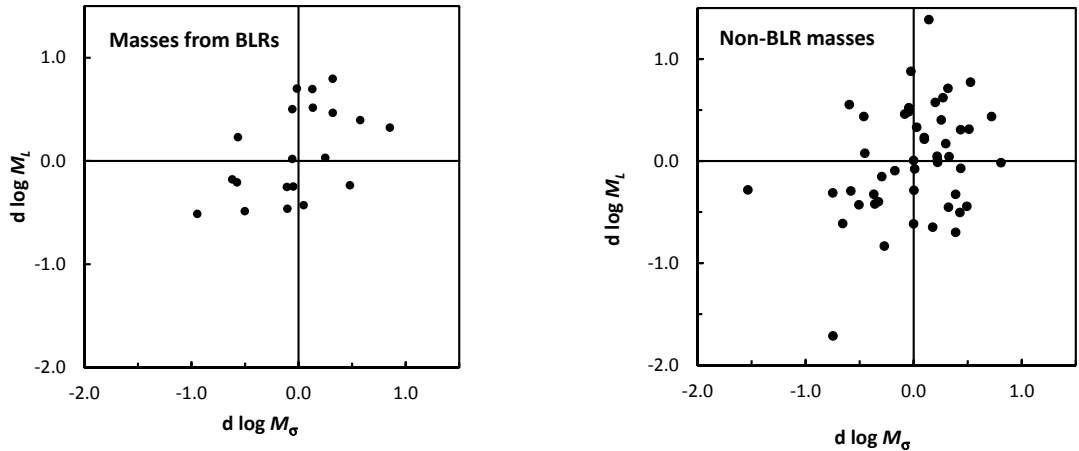}
%% Note the ABSENCE of the extension .pdf , .eps or .ps  !
  \caption{ $d\log M_L$, the residuals from the $M_\bullet$ -- $L_{bulge}$ relationship, versus $d\log M_\sigma$, the residuals from the $M_\bullet$ -- $\sigma_*$ relationships.   The left frame is for the $M_{new}$ estimates for the AGNs discussed here and the right frame shows data from \citet{gultekin+09}. }
  \label{Gaskell:Fig3}
\end{figure}

Since $M_\bullet$ appears on both axes in Fig.\@ 3, if there are large errors in the mass determinations they will introduce a direct correlation into the residuals.  No strong correlations can be seen.  The slight elongation in Fig.\@ 3a corresponds to possible scatter in $M_{new}$ of about 0.28 dex.

In Fig.\@ 2b it can be seen that the scatter in the $M_\bullet$ -- $L_{bulge}$ relationship declines with increasing mass.  Fig.\@ 4 shows the scatter in the $M_\bullet$ -- $L_{host}$ relationship as a function of $L_{host}$.

There are several conclusions that can be drawn from Fig.\@ 4.
\begin{enumerate}
\item The scatter for the highest luminosity bin ($\pm 0.14$ dex) is remarkably small.  It is comparable to the \cite{denney+09} estimate of the measuring error. This thus provides good support for the accuracy of the approach proposed in \citet{gaskell10a}.
\item Since the $M_{new}$ determinations seem accurate, the substantial scatter at low luminosities must be {\em intrinsic}.
    \end{enumerate}

It is possible that there is some unknown additional source of error causing the increase in dispersion in Fig.\@ 4 at low masses, but a similar trend has already been reported for a different sample for the dispersion in the $M_\bullet$ -- $\sigma_*$ relationship \citep{gaskell09a}.  Furthermore, the wide dispersion in $M_\bullet$ found from highly-accurate maser measurements of yet another sample \citep{greene+10} provides additional strong independent support for the large increase in scatter in $M_\bullet$/$L_{host}$ at low masses being real.

\vspace{0.1 cm}

%% Figure 4
\begin{figure}[ht!]
 \centering
 \includegraphics[width=0.4\textwidth,clip]{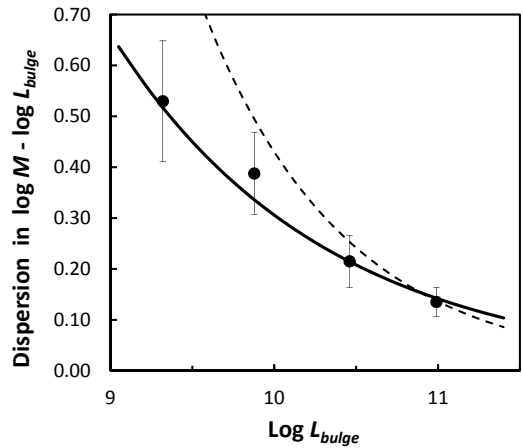}
%% Note the ABSENCE of the extension .pdf , .eps or .ps  !
  \caption{The scatter in the AGN $M_\bullet$ -- $L_{host}$ relationship as a function of bulge luminosity.  The AGNs have been divided equally into four $L_{host}$ bins.  The dotted line shows the expected $n^{-1/2}$ reduction in the scatter if a galaxy forms as a result of mergers of $n$ units with Gaussian distributions of $M_\bullet$ and $L_{stars}$.  The solid line shows the result of simulations using a log-normal initial distribution of black hole mass to stellar mass ratio.  (Figure adapted from \citealt{gaskell10a})}
  \label{Gaskell:Fig4}
\end{figure}

\section{The origin of $M_\bullet$ -- $L_{host}$ relationship}

It has long been recognized that mergers will influence correlations between $M_\bullet$ and bulge properties.  For example, \citet{haehnelt+kauffmann00} pointed out that mergers move galaxies along the bulge/SMBH scaling relationships.  \citet{peng07} discussed how mergers might affect the bulge/SMBH scaling relationships and made the important point that SMBH--bulge scaling relationships will emerge, independent of the initial conditions and detailed physics, simply because of averaging and that ``random merging of galaxies that harbor random black hole masses tends to strengthen rather than weaken a preexisting, linear, correlation.'' \citet{gaskell10c} provided the first observational evidence that merging alone is indeed the origin of the $M_\bullet$ -- $L_{host}$ relationship and further proposed that a preexisting correlation is neither necessary nor desirable. If we assume that initially $M_\bullet$ and the stellar luminosity of each galactic building block have independent normal distributions, then, if we merge $n$ building blocks, $L_{bulge}$ is proportional to $n$ and the scatter in the $M_\bullet$/$M_{bulge}$ ratio goes down as $n^{-1/2} = L_{bulge}^{-1/2}$.  This is shown in Fig.\@ 4 as the dotted line.  This is probably too steep a decrease in the scatter.\footnote{I say ``probably'' because the right-most point must be somewhat too high because of measuring error, and selection effects (not observing very weak AGNs with small SMBHs) could well be lowering the dispersion at low luminosities.}  However, a $n^{-1/2}$ decrease is only true for a normal distribution of initial values of $M_\bullet$ and stellar luminosity.  The actual distributions are almost certainly highly non-Gaussian.  Simulations were therefore run starting with a more likely log--normal distribution.  These gives the solid curve in Fig.\@ 4.

Several important conclusions follow from these results:
\begin{enumerate}
\item The $M_\bullet$ measurements {\em require} a large scatter in $M_\bullet$/$L_{stars}$ in the initial merging units --- a scatter greater than that for the least massive galaxies in the sample, where the $\pm$2$\sigma$ scatter is already a factor of $\sim 300$.
\item Because of this, detectable SMBHs in {\it Sc}/{\it Sd} galaxies are going to be rare.
\item Intermediate mass black holes (IMBHs) in globular clusters will also be very rare.
\item As pointed out in \citet{gaskell10c}, the huge scatter in $M_\bullet$/$L_{stars}$ in the initial merging units {\em excludes} fine-tuning of $M_\bullet$ or $L_{bulge}$ or else the subsequent scatter in larger galaxies would be much lower than observed.
\item Fine-tuning via AGN feedback that significantly reduces scatter at any subsequent stage is also excluded.  This is not saying, of course, that an AGN has zero effect on its host galaxy --- only that it does not {\em significantly} affect SMBH--bulge relationships.
\item The averaging due to mergers will only produce an SMBH--bulge relationship in {\em classical bulges} --- not in psuedo-bulges which grow through secular processes.  This has indeed been found to be the case \citep{kormendy+11}.
\item The model explains why $M_\bullet$ is systematically higher at a given $\sigma_*$ in classical bulges than in pseudo bulges \citep{graham08,hu08,graham+11}, because, in classical bulges, $M_\bullet$ is dominated by the highest $M_\bullet$ in last merger.
\end{enumerate}

The simulations discussed in \citet{gaskell10c} and here are purely for ``dry'' mergers (those without star formation), but SMBHs certainly grow through the accretion of gas as do the stellar populations of galaxies. I believe that including star-forming ``wet'' mergers will not substantially alter the conclusions for two reasons: firstly, at the high-mass end we are dealing with galaxies becoming ``red and dead''.  Mergers between them will effectively be dry.  The second reason why wet mergers will not have a large effect on Fig.\@ 4 is that they are most important at the low mass end where the scatter is already large.  There are two main possibilities here: either the stellar population and the SMBH grow roughly proportionately or they do not.  The former gives the same result as dry mergers, while the latter can increase the dispersion in the $M_\bullet$/$L_{stars}$ ratio.  However, as mentioned, this dispersion is already very large, so the effect on Fig.\@ 4 will be negligible.  \citet{jahnke+maccio11} have recently performed more elaborate simulations including prescriptions for star formation, black hole growth, and disk-to-bulge conversion, and they have affirmed the conclusions of \citet{gaskell10c}.

\section{Conclusions}

A slight modification to the Dibai method seems to significantly increase the accuracy of BH mass estimates. Using these new mass estimates the scatters about the $M_\bullet$ -- $L_{bulge}$ and $M_\bullet$ -- $\sigma_*$ relationships are comparable (contrary to what was thought when the $M_\bullet$ -- $\sigma_*$ relationship was discovered).
These mass estimates imply that there is a strong decrease in the dispersion of the $M_\bullet$ -- $L_{host}$ relationship for higher mass black holes and hosts.  This is qualitatively consistent with the BH--bulge relationships being {\em solely a consequence of mergers and not having an underlying physical cause.}  The dispersion at the low mass end is too high to allow any significant causal relationship between black hole and bulge masses.

% Optional acknowledgements
% -------------------------
\begin{acknowledgements}
I am grateful to John Kormendy and Jenny Greene for useful discussions.  This research has been supported by US National Science Foundation grants AST 03-07912 and AST 08-03883, NASA grant NNH-08CC03C, and grant 32070017 of the GEMINI-CONICYT Fund.
\end{acknowledgements}

%%-----------------------------
%%   Bibliography
%%-----------------------------
%%
%% The reference list should contain all the references cited in the text, ordered alphabetically by surname (with
%% initials following). If there are several references to the same first author, they should be entered according
%% to the following scheme:
%% 1. One author: chronologically
%% 2. Author, one co-author: alphabetically by co-author, then chronologically
%% 3. Author, two or more co-authors: chronologically.
%%
%% Please note that for papers that have more than five authors, only the first three should be given, followed
%% by "et al."
%%
%% The format for references is the one adopted by A&A (see the example below).
%%
%% To set the reference list in the proper A&A format, we encourage you to use BibTEX and the natbib
%% package instead of the standard 'thebibliography' environment.
%%

%% The following lines are required when using BibTEX (strongly encouraged!):
\bibliographystyle{aa}  % A&A bibliography style file (aa.bst)
\bibliography{sf2a-template} % your references in file: Yourfile.bib

\end{document}